\documentclass[prl,aps,amsmath,amsfonts,amssymb,twocolumn,superscriptaddress,footinbib,showpacs,10pt]{revtex4}
\usepackage{times,xspace}
\usepackage{amsfonts}
\usepackage{bm}
\usepackage{amssymb}
\usepackage[final]{graphicx}
\usepackage{color}
\begin{document}
\title{Interaction-induced merging of Landau levels in an electron system of double quantum wells}
\author{A.~A. Shashkin}
\affiliation{Institute of Solid State Physics, Chernogolovka, Moscow District, 142432,Russia}
\author{V.~T. Dolgopolov}
\affiliation{Institute of Solid State Physics, Chernogolovka, Moscow District, 142432,Russia}
\author{J.~W. Clark}
\affiliation{McDonnell Center for the Space Science and Department of Physics, Washington University, St. Louis, MO 63130, USA}
\author{V.~R. Shaginyan}
\affiliation{Petersburg Nuclear Physics Institute, NRC Kurchatov Institute, Gatchina, 188300, Russia}
\author{M.~V. Zverev}
\affiliation{National Research Centre Kurchatov Institute, Moscow, 123182, Russia}
\affiliation{Moscow Institute of Physics and Technology, Dolgoprudny, Moscow District, 141700, Russia}
\author{V.~A. Khodel}
\affiliation{National Research Centre Kurchatov Institute, Moscow, 123182, Russia}
\affiliation{McDonnell Center for the Space Science and Department of Physics, Washington University, St.
Louis, MO 63130, USA}

\begin{abstract}
We show that the disappearance of the chemical potential jumps over
the range of perpendicular magnetic fields at fixed integer filling
factor in a double quantum well with a tunnel barrier is caused by
the interaction-induced level merging. The distribution function in
the merging regime is special in that the probability to find an
electron with energy equal to the chemical potential is different
for the two merged levels.
\end{abstract}

\maketitle

More than twenty years ago a topological phase transition that is
related to the emergence of a flat portion of the single-particle
spectrum $\varepsilon(k) $ at the chemical potential was predicted
at $T=0$ in strongly correlated Fermi systems
\cite{khodel90,volovik91,nozieres92,khodel08,yudin14}. In more vivid
terms, this transition is associated with the band flattening or
swelling of the Fermi surface (for recent reviews, see
Refs.~\cite{amusia14,clark12,zverev12}). The flattening of the
single-particle spectrum means that the probability to find a fermion
with energy equal to the chemical potential depends on the fermion
momentum $k$. The swelling of the Fermi surface is preceded by an
unleashed increase of the quasiparticle effective mass $m$ at the
quantum critical point \cite{shashkin02,dolgopolov15}.

The topological phase transition characterized by the unusual form
of the distribution function is not the only non-trivial
manifestation of fermion interactions in strongly correlated Fermi
liquids. Another example is the merging of quantum levels in a Fermi
system with discrete spectrum in which case the fillings of the two
quantum levels at the chemical potential are different
\cite{khodel07}. The merging of the spin- and valley-split Landau
levels at the chemical potential has been detected near the quantum
critical point in a clean strongly-interacting two-dimensional (2D)
electron system in (100) Si metal-oxide-semiconductor field-effect
transistors (MOSFETs) \cite{shashkin14}. The fact that the merging
detected is governed by the effective mass depending on electron
density may create the impression that the level merging is a
precursor of the swelling of the Fermi surface. As a matter of fact,
the two effects are not always related to each other. The diverging
effective mass is not necessary for the existence of the effect of
level merging.

Here, we show that the disappearance of the chemical potential jumps
over the range of perpendicular magnetic fields at fixed integer
filling factor in a double quantum well with a tunnel barrier is
caused by the interaction-induced level merging. The merging regime
corresponds to the special form of the distribution function. In
this case the probability to find an electron with energy equal to
the chemical potential is different for the two merged levels.

\begin{figure}
\scalebox{.45}{\includegraphics{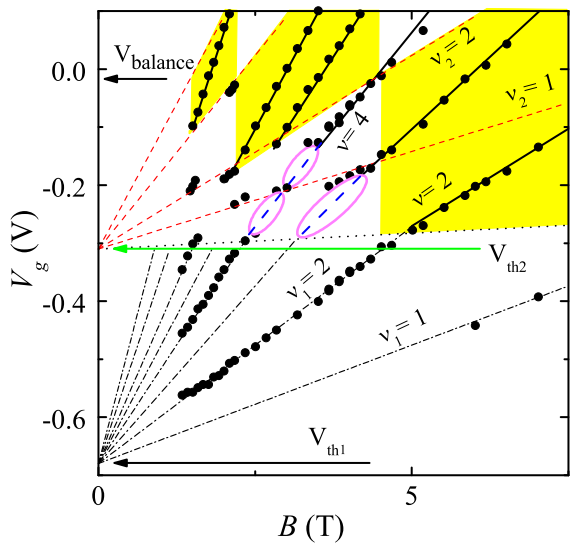}}
\caption{\label{fig1} Landau level fan chart for the double quantum
well shown in Fig.~\ref{fig2}. Positions of the longitudinal
resistance minima in the $(B,V_g)$ plane are marked by the dots. The
filling factor $\nu$ for the double layer electron system as well as
the filling factor $\nu_1$ ($\nu_2$) for the back (front) layer are
indicated. Over the shaded areas, the merging of quantum levels in
perpendicular magnetic fields is impossible. In the regions marked by
the ovals, no resistance minima are observed in a perpendicular
magnetic field, whereas these appear in a tilted magnetic field.}
\end{figure}

To begin with, we assume that the weakly interacting two-dimensional
electron system is subjected to a perpendicular magnetic field $B$.
In the simplest case there are two equidistant ladders of quantum
levels for spin up and down directions (see, e.g.,
Ref.~\cite{dolgopolov14}). Both the thermodynamic and kinetic
properties of the electron system are determined by the position of
the chemical potential relative to the quantum levels, which is in
turn determined by the magnetic field and electron density $n$. The
filling factor is equal to $\nu=n/n_0$, where $n_0=eB/hc$ is the
level degeneracy. When $\nu$ is fractional, the chemical potential is
pinned to the partially filled quantum level. The probability to find
an electron at the chemical potential is given by the fractional part
of the filling factor and can be varied between zero and one. At
integer filling factor there is a jump of the chemical potential. In
experiment, the jump manifests itself as a minimum in the
longitudinal electrical resistance in the Shubnikov-de Haas effect.
The resistance minima in the $(B,n)$ plane correspond to a Landau
level fan chart like the one shown in Fig.~\ref{fig1}.

If the magnetic field is tilted by an angle $\beta$, the spacing
between the quantum levels in each of the spin ladders is equal to
$\hbar\omega_c=\hbar eB\cos(\beta)/mc$, and the shift between the
ladders equals $g\mu_BB$, where the Lande factor $g$ is assumed to
be isotropic, $\mu_B=e\hbar/2m_ec$ is the Bohr magneton, and $m_e$ is
the free electron mass. Increasing the tilt angle leads to crossing
the quantum levels of the two ladders. The crossing happens for the
first time at an angle $\beta_1$ that satisfies the condition

\begin{equation}
\cos(\beta_1)=\frac{gm}{2m_e}.\label{eq0}
\end{equation}
At $\beta=\beta_1$, the chemical potential jumps at even filling
factors and the corresponding fan chart lines will disappear. This
effect is well known and is used for the experimental determination
of the product $gm$ \cite{ando82,kravchenko00,pudalov02}. Note that
in experiment, the chemical potential jumps should be absent at tilt
angles in the vicinity of $\beta_1$, depending on the sample quality
and temperature.

We now take into account the interaction between the electrons of
neighboring quantum levels and increase the tilt angle in the
vicinity of $\beta_1$. Tentatively, the quantum level filled before
crossing should have got emptied with increasing $\beta$. However,
if the single-particle energy of electrons on the emptying level
decreases due to the electron interaction, both levels remain pinned
to the chemical potential over a wide range of angles
$\Delta\beta_1$ that is determined by the interaction strength. The
probability to find an electron at the chemical potential is
different for opposite spin orientations, being dependent on the
external parameter which is the tilt angle. Such a behavior is known
as the merging of quantum levels.

In the above hypothetical consideration the crossing or merging of
quantum levels is controlled by the tilt angle of the magnetic
field. In the experiments on a strongly-interacting 2D electron
system in (100) Si MOSFETs, the disappearance of the longitudinal
resistance minima is analyzed when changing both the perpendicular
magnetic field and electron density at fixed filling factor
$\nu=4(i+1)$, where $i$ is an integer. In this case the level
merging occurs near the quantum critical point, as controlled by the
effective mass depending on electron density \cite{shashkin14}. One
might think that the level merging is a precursor of the Fermi
surface swelling. In fact, the two effects are not necessarily
related to each other. Below, we demonstrate that the effect of level
merging occurs in a bilayer 2D electron system with a tunnel barrier
between the electron layers. Note that although the Shubnikov-de Haas
effect in similar double layer electron systems was investigated in a
number of publications
\cite{davies96,dolgopolov99,deviatov00,zhang07,duarte07}, only the
level crossing was observed in
Refs.~\cite{davies96,zhang07,duarte07}.

\begin{figure}
\scalebox{.42}{\includegraphics{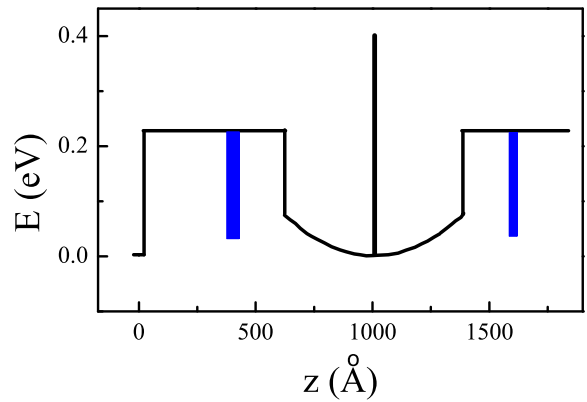}}
\caption{\label{fig2} Schematic diagram of the bottom of the
conduction band for the AlGaAs double quantum well in the absence of
electrons. The parabolic part of the well is grown when varying the
Al content from zero at the center to 0.1 on the edge of the well.
The tunnel barrier at the center is created by three monolayers of
Al$_x$Ga$_{1-x}$As ($x=0.3$). The thick blue lines correspond to the
silicon doped layers.}
\end{figure}

\begin{figure}
\scalebox{.42}{\includegraphics{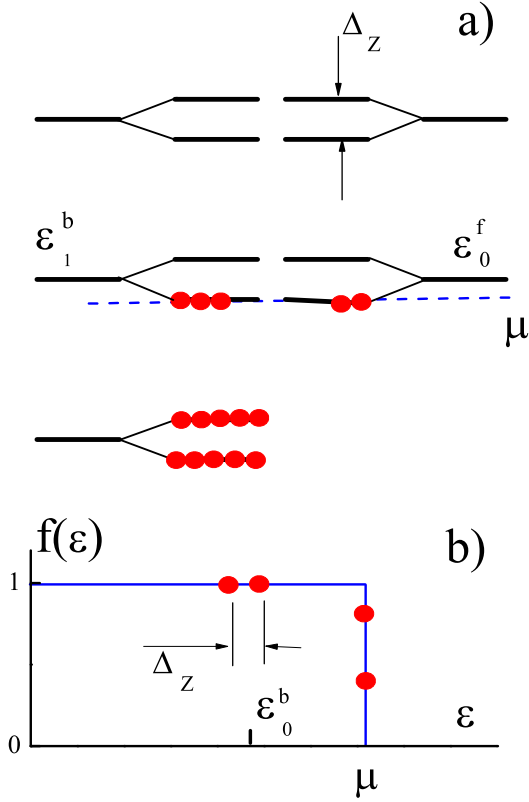}}
\caption{\label{fig3} (a) Layout and filling of the quantum levels in
the bilayer electron system in the merging regime at filling factor
$\nu=3$. (b) The distribution function of the electrons in the
merging regime at $\nu=3$.}
\end{figure}

The samples used contain a parabolic quantum well grown on a GaAs
substrate, as shown schematically in Fig.~\ref{fig2}. The width of
the parabolic part of the well, limited by vertical walls, is about
760~\AA . At the center of the well there is a narrow tunnel barrier
that consists of three monolayers of Al$_x$Ga$_{1-x}$As ($x=0.3$).
The symmetrically doped structure is capped by 600~\AA\ AlGaAs and
40~\AA\ GaAs layers over which a metallic gate is evaporated. The
presence of the tunnel barrier leads to a splitting of each subband
bottom caused by quantization in the $z$ direction. At the point of
the symmetric electron density distribution, the splitting energy is
equal to 1.3~meV. The structure of the quantum levels in the bilayer
2D electron system in perpendicular magnetic fields is similar to
that in the 2D electron system in (100) Si MOSFETs, where the spin
and valley splittings are present, with a distinction that the spin
splitting $\Delta_Z$ in accessible magnetic fields is the smallest
(Fig.~\ref{fig3}(a)).

Applying a voltage $V_g$ between the gate and the contact to the
quantum well makes it possible to tune the electron density. The
electrons appear in the back part of the quantum well when the gate
voltage is above $V_{th1}\approx -0.7$~V and occupy one subband up
to $V_g=V_{th2}\approx -0.3$~V (Fig.~\ref{fig1}). At $V_g> V_{th2}$,
the electrons appear in the front part of the well and fill the
second subband up to the balance point $V_g=V_{balance}\approx 0$.
This behavior is typical of the so-called ``soft'' double layer
electron systems \cite{dolgopolov99,deviatov00,gold94}.

We focus on the range of gate voltages between $V_{th2}$ and
$V_{balance}$ where the electrons in perpendicular magnetic fields
occupy the two quantum ladders. Positions of the quantum levels are
determined by the magnetic field and gate voltage. Over the shaded
areas in Fig.~\ref{fig1}, the gaps in the single-particle spectrum
and the chemical potential jumps are protected by quantum effects
\cite{dolgopolov99}. In the remaining areas, the merging of quantum
levels is in principle possible.

We now consider the filling factor $\nu=3$. At $V_g=V_{th2}$, the
magnetic field is equal to $B_\nu=n_1(V_{th2})hc/3e$, where
$n_1(V_{th2})$ is the electron density in the back layer. The energy
$\varepsilon_0-g\mu_BB_\nu/2$ of the spin up level of the front layer
ladder is the same as the energy $\varepsilon_1-g\mu_BB_\nu/2$ of the
spin up level of the back layer ladder (Fig.~\ref{fig3}(a)). Since
far from the balance point, the electron density in the back layer
remains practically unchanged with increasing $V_g$ above $V_{th2}$
(see, e.g., Fig.~1(b) of Ref.~\cite{davies96}), the electron density
in the front layer in a magnetic field $B=B_\nu+\Delta B$ along the
dashed line at $\nu=3$ restricted by the oval in Fig.~\ref{fig1} is
equal to $n_2\simeq\Delta n=3e\Delta B/hc$. To balance the change in
the cyclotron energy $\hbar e\Delta B/mc$ and have both levels pinned
to the chemical potential $\mu$, it is necessary to transfer a small
amount of electrons between the levels

\begin{equation}
\delta n_2=-\delta n_1>0,\label{eq1}
\end{equation}
which gives rise to a shift of the single-particle levels

\begin{eqnarray}
\delta\varepsilon_1^b&=&\Gamma_1^{bb}\delta n_1+\Gamma_1^{bf}\delta n_2 \nonumber\\
\delta\varepsilon_0^f&=&\Gamma_0^{fb}\delta n_1+\Gamma_0^{ff}\delta n_2,\label{eq2}
\end{eqnarray}
where $\Gamma_j^{\lambda\sigma}$ is the electron interaction
amplitude, the index $b$ ($f$) refers to the back (front) layer, and
$j=0,1$ is the Landau level number. Both levels are pinned to the
chemical potential under the condition
$\delta\varepsilon_0-\delta\varepsilon_1=\hbar e\Delta B/mc$, which
yields

\begin{equation}
\delta n_2=\frac{\hbar e\Delta B}{mc\Gamma},\label{eq3}
\end{equation}
where $\Gamma=\Gamma_1^{bb}+\Gamma_0^{ff}-\Gamma_1^{bf}-
\Gamma_0^{fb}$. In the plate-capacitor approximation, we get

\begin{equation}
\Gamma\simeq\frac{4\pi e^2a}{\epsilon},\label{eq4}
\end{equation}
where $a$ is the distance between the weight centers of the electron
density distributions in the $z$ direction in both subbands. The
level merging holds for the filling factor

\begin{equation}
\nu_2=\frac{(n_2+\delta n_2)hc}{eB}<1.\label{eq5}
\end{equation}

We stress that in the wide range of magnetic fields at fixed filling
factor $\nu$, the probability to find an electron with energy equal
to the chemical potential is different for the two merged levels, as
shown in Fig.~\ref{fig3}(b).

It is easy to verify that the pinning of the levels
$\varepsilon_0^f$ and $\varepsilon_1^b$ to the chemical potential is
energetically favorable. The variation of the system energy $\delta
E$ caused by electron density redistribution is given by the Fermi
liquid relation

\begin{eqnarray}
\delta E=(\varepsilon_1^b-g\mu_BB/2)\delta n_1+(\varepsilon_0^f-g\mu_BB/2)\delta n_2 \nonumber\\
+\frac{1}{2}(\Gamma_1^{bb}\delta n_1^2+\Gamma_0^{ff}\delta n_2^2+\Gamma_1^{bf}\delta n_1\delta n_2+\Gamma_0^{fb}\delta n_2\delta n_1).\label{varener1}
\end{eqnarray}
Using Eq.~(\ref{eq1}), one gets

\begin{equation}
\delta E=-\frac{\hbar e\Delta B}{mc}\delta n_2+\frac{1}{2}\Gamma\delta n_2^2.\label{varener2}
\end{equation}
The redistribution (\ref{eq3}) yields the negative shift in the
energy of the system

\begin{equation}
\delta E=-\frac{1}{2}\frac{\hbar e\Delta B}{mc}\delta n_2.\label{varener3}
\end{equation}

Let the variation $\delta' n_2$ be different from $\delta n_2$ of
Eq.~(\ref{eq3}). Substituting $\delta' n_2$ for $\delta n_2$ in
Eq.~(\ref{varener2}), one obtains the energy shift

\begin{equation}
\delta' E=\delta E+\frac{1}{2}\Gamma(\delta' n_2-\delta n_2)^2.\label{varener4}
\end{equation}
Obviously, the redistribution (\ref{eq3}) corresponding to the
merging of the levels $\varepsilon_0^f$ and $\varepsilon_1^b$
provides the maximum gain in the system energy.

For the sake of simplicity, we have used in our argumentation the
case of filling factor $\nu=3$. Still, the same arguments are valid
for higher filling factors.

The occurrence of the merging of quantum levels in the experiment is
confirmed by using tilted magnetic fields. With tilting magnetic
field the magnetoresistance minima and chemical potential jumps
arise \cite{deviatov00} particularly along the dashed lines at
$\nu=3$ and $\nu=4$ indicated by the ovals in Fig.~\ref{fig1}. The
appearance of the chemical potential jumps in the double layer
electron system in tilted magnetic fields signals that the quantum
levels are narrow enough.

As has been mentioned above, the chemical potential jumps can be
protected by quantum effects. In general, a transfer of electrons
between the quantum levels of different subbands leads to mixing the
wave functions of the subbands and opening an energy gap if the
non-diagonal matrix elements are not equal to zero
\cite{dolgopolov99}. This is realized over the shaded areas in
Fig.~\ref{fig1}. In contrast, in the merging regions at $\nu=3$ and
$\nu=4$ indicated by the ovals in Fig.~\ref{fig1}, the non-diagonal
matrix elements in perpendicular magnetic fields equal zero because
of orthogonality of the in-plane part of the wave functions in the
bilayer electron system. Tilting the magnetic field breaks the
orthogonality of the wave functions of the neighboring quantum
levels, and the energy gap emerges \cite{deviatov00,duarte07}.

In a number of publications \cite{davies96,zhang07,duarte07}, the
so-called ring-like structures were revealed by transport
measurements in double layer or double subband electron systems. The
ring-like structures corresponding to the maximum resistance are
topologically equivalent to the regions in between the data points in
Fig.~\ref{fig1} with a distinction that at fixed integer filling 
factor $\nu$, the latter regions are considerably wider, as expected.

It is easy to formulate the condition when the merging of quantum
levels is observed. For the level merging to occur in a wide range of
magnetic fields, the value $\delta n_2$ of Eq.~(\ref{eq3}) must be
small compared to $n_2$, which gives an estimate for $\Gamma$

\begin{equation}
\Gamma\gg\frac{2\pi\hbar^2}{3m}.\label{eq6}
\end{equation}
Using the evaluation of Eq.~(\ref{eq4}), we arrive at the condition
of the level merging

\begin{equation}
\alpha=\frac{a}{a_B}\gg 1,\label{eq7}
\end{equation}
where $a_B$ is the effective Bohr radius. It is the relation that
describes the softness of a bilayer electron system, i.e., the
sensitivity of the subband spacing to intersubband electron transfer.

It is worth noting that in experiments, the quantum level width and
temperature are always finite. The merging at finite temperatures
implies that the single-particle energies of the merged levels have
an energy difference $\lesssim k_BT$ so that the levels converge with
decreasing temperature. On the other hand, the value $\delta n_2$ is
restricted from below: $S\delta n_2\geq 1$, where $S$ is the sample
area. The finite value of $\delta n_2$ means that the energy
difference between the levels at filling factor $\nu=3$ exceeds
$\Gamma/S$. This determines the characteristic temperature

\begin{equation}
T^*\simeq\frac{\Gamma}{k_BS}.\label{eq8}
\end{equation}
Using the realistic values $a=300$~\AA\ and $S=10\times
10$~$\mu$m$^2$, one estimates $T^*\sim 5$~mK. While at $T\gg T^*$ the
electron system with the merged levels is characterized by the
conventional Fermi distribution, in the opposite limit $T\ll T^*$
(and the small level width compared to $k_BT^*$) the distribution
function has non-Fermi liquid form.

For the non-Fermi distribution function, the finite entropy
persisting down to very low temperatures threatens violation of the
Nernst theorem. Conventionally, it is supposed that as $T\to 0$,
shedding the paradoxical entropy excess occurs by virtue of one of
the possible mechanisms discussed in Ref.~\cite{clark12}. For
example, in case of Cooper-pair formation, the gap $\Delta(T)$ in the
spectrum ensures vanishing entropy $S(T\to
0)\propto\exp(-\Delta(0)/k_BT)$. Assuming that the magnetic field is
weak enough to save the superconductivity, the critical temperature
$T_c$ of the possible Cooper transition for our case can be evaluated
at $T_c\sim T^*/\ln(\alpha)$.

In summary, we have shown that the merging of quantum levels is not
necessarily a precursor of the Fermi surface swelling. We demonstrate
the occurrence of the level merging in a soft double layer electron
system in perpendicular magnetic fields. The distribution function in
the merging regime has non-Fermi liquid form when the probability to
find an electron with energy equal to the chemical potential is
different for the two merged levels.

This work was supported in part by RFBR 15-02-03537, 13-02-00095,
15-02-06261, and 14-02-00044, RAS, the Russian Ministry of Sciences,
NS-932.2014.2, RCSF 14-12-00450, the U.S. DOE, Division of Chemical
Sciences, the Office of Basic Energy Sciences, the Office of Energy
Research, AFOSR, and the McDonnell Center for the Space Sciences.

\end{document}